\documentclass[ twocolumn]{aastex6}
\usepackage{times}
\usepackage{graphicx}

\def \aj {AJ}
\def \mnras {MNRAS}
\def \pasp {PASP}
\def \apj {ApJ}
\def \apjs {ApJS}
\def \apjl {ApJL}
\def \aap {A\&A}
\def \nat {Nature}
\def \araa {ARAA}

\def\lesssim{\mathrel{\hbox{\rlap{\hbox{\lower4pt\hbox{$\sim$}}}\hbox{$<$}}}}
\def\gtrsim{\mathrel{\hbox{\rlap{\hbox{\lower4pt\hbox{$\sim$}}}\hbox{$>$}}}}

\long\def\symbolfootnote[#1]#2{\begingroup%
\def\thefootnote{\fnsymbol{footnote}}\footnote[#1]{#2}\endgroup} 
\watermark{}
\begin{document}
\shorttitle{SN 2006jc companion}
\shortauthors{Maund et al.}
\title{The possible detection of a binary companion to a Type Ibn supernova progenitor}
\author{J.R. Maund\altaffilmark{1,2}, A. Pastorello\altaffilmark{3},  S. Mattila\altaffilmark{4, 5}, K. Itagaki\altaffilmark{6} and T. Boles\altaffilmark{7}}
\altaffiltext{1}{Department of Physics and Astronomy, University of Sheffield, Hicks Building, Hounsfield Road, Sheffield, S3 7RH, U.K.}
\altaffiltext{2}{Royal Society Research Fellow}
\altaffiltext{3}{INAF - Osservatorio Astronomico di Padova. Vicolo Osservatorio 5, I-35122, Padova, Italy}
\altaffiltext{4}{Tuorla Observatory, Department of Physics and Astronomy, University of Turku, V\"ais\"al\"antie 20, FI-21500 Piikki\"o, Finland}
\altaffiltext{5}{Finnish Centre for Astronomy with ESO (FINCA), University of Turku, V\"ais\"al\"antie 20, FI-21500 Piikki\"o, Finland}
\altaffiltext{6}{Itagaki Astronomical Observatory, Teppo-cho, 990-2492 Yamagata, Japan}
\altaffiltext{7}{Coddenham Astronomical Observatory, Suffolk IP6 9QY, UK}
\begin{abstract}
We present late-time observations of the site of the Type Ibn supernova (SN) 2006jc acquired with the Hubble Space Telescope Advanced Camera for Surveys.  A faint blue source is recovered at the SN position with brightness $m_{F435W}=26.76\pm0.20$, $m_{F555W}=26.60\pm0.23$ and $m_{F625W} = 26.32\pm0.19$ mags, although there is no detection in a contemporaneous narrow-band $\mathrm{H\alpha}$ image.  The spectral energy distribution of the late-time source is well fit by a stellar-like spectrum ($\log T_{eff} > 3.7$ and $\log L / L_{\odot} > 4$) subject to only a small degree of reddening consistent with that estimated for SN~2006jc itself at early-times.  The lack of further outbursts after the explosion of SN~2006jc suggests that the precursor outburst originated from the progenitor. The possibility of the source being a compact host cluster is ruled out on the basis of the source's faintness, however the possibility that the late-time source maybe an unresolved light echo originating in a shell or sphere of pre-SN dust (within a radius $1\mathrm{pc}$)  is also discussed.  Irrespective of the nature of the late-time source, these observations rule out a luminous blue variable as a companion to the progenitor of SN~2006jc.
\end{abstract}

\maketitle
\section{Introduction}
\label{intro}
Supernova (SN) 2006jc is the prototype of the class of hydrogen-deficient SNe that exhibit prominent narrow lines of He~I in their spectra, the  so-called ``Type Ibn" SNe \citep{2008MNRAS.389..113P,2016MNRAS.456..853P}.  SN~2006jc also has the distinction of being the first SN for which a pre-explosion outburst was detected; in the case of SN~2006jc an outburst was previously identified at the SN position in 2004, hence 2 years prior to the SN itself \citep{2006CBET..666....1N,2007Natur.447..829P,2007ApJ...657L.105F}.  Outbursts of this sort are usually associated with Luminous Blue Variables (LBVs) for which eruptions are considered a natural part of their evolution, despite  the exact physics responsible for them is poorly known \citep{2014ARA&A..52..487S}. \citet{galyam05gl} first confirmed the presence of a bright, massive ($\sim 60M_{\odot}$) LBV-like progenitor in pre-explosion observations of the Type IIn SN 2005gl.  LBVs provide a natural connection between pre-explosion outbursts and the creation of dense H-rich circum-stellar envelopes with which subsequent SNe interact yielding Type IIn SNe.  In contrast, Type Ibn SNe require the presence of an He-rich and, generally, H-poor circumstellar medium (CSM) arising from a Wolf-Rayet (WR) progenitor undergoing an LBV-like eruption,  although this interpretation has been somewhat questioned with the discovery of a Type Ibn SN in the outskirts of an elliptical galaxy \citep[hence in a likely old stellar population environment;][]{2013ApJ...769...39S}.

So far, about two dozen Type Ibn SNe have been discovered  \citep[see, e.g., the samples of][]{2008MNRAS.389..113P,2016MNRAS.456..853P,2016AAS...22712003H}, and no other example has shown an LBV-like outburst similar to that observed before the explosion of SN~2006jc. In addition, no progenitor of a Type Ibn SN has ever been seen in quiescence \citep{2014ARA&A..52..487S} to unequivocally prove the WR nature of their progenitors. For all these reasons, the stellar configuration that produced the sequence of events observed at the location of SN~2006jc is still debated.

The most common interpretation for SN~2006jc and its precursor outburst, is that the progenitor was a WR star with residual LBV-like instability \citep[e.g.][]{2007Natur.447..829P,2007ApJ...657L.105F,2008ApJ...687.1208T}.  An alternative scenario, however, was proposed by \citet{2007Natur.447..829P} for SN~2006jc, invoking a massive binary system to explain the chain of events that occurred in 2004-2006: an LBV companion erupted in 2004, and while it was a WR star that exploded in 2006 as a normal stripped-envelope SN. Nonetheless, this scenario did not comfortably explain the narrow He~I emission features observed in the spectrum of SN~2006jc.  Furthermore, if SN~2006jc occurred in a massive star forming region there was the possibility that the precursor outburst may have arisen from an LBV-like star in close proximity to the progenitor but otherwise unrelated to the explosion.

Here we present late-time observations of the site of SN~2006jc, to explore the nature of the progenitor systems responsible for Type Ibn SNe. 
SN~2006jc occurred in UGC 4904\footnote{https://ned.ipac.caltech.edu}, for which the corrected recessional velocity is $2029\pm19\,\mathrm{km\,s^{-1}}$.  Assuming $H_{0} = 73\,\mathrm{km\,s^{-1}\,Mpc^{-1}}$, we adopt a distance of $27.8\pm1.9\,\mathrm{Mpc}$.   SN 2006jc was discovered by K. Itagaki \citep{2006CBET..666....1N} on 2006 Oct 9, and analysis by \citet{2007Natur.447..829P} suggested it was discovered only a few days after maximum.  More recently an earlier observation of UGC~4904 acquired by K. Itagaki on 2006 Oct 3 was found, in which SN 2006jc was detected at $m_{R}= 13.44\pm0.27\,\mathrm{mags}$.    Comparisons of the lightcurve of SN~2006jc with other SNe with similar decline rates suggests that SN 2006jc may have exploded on the order of $\sim 8$ days before maximum \citep{2007Natur.447..829P,2007ApJ...657L.105F}.  We assume a metallicity for the site of SN~2006jc of half-solar, or Large Magellanic Cloud, abundances following the measurement of \citet{2015AA...580A.131T}.

\section{Observations}
\label{sec:obs}

\begin{table}
\caption{\label{tab:obs} HST observations of the site of SN~2006jc}
\begin{tabular}{lcccc}
\hline\hline
Date     & Instrument & Filter  & Exposure & Program \\
(UT)      &                   &           & Time (s)   &               \\
\hline
2008 Nov 19.9 & $WFPC2/PC$ & $F555W$ & 460 & 10877$^{1}$\\
2008 Nov 19.9 & $WFPC2/PC$ & $F814W$ & 700 & 10877\\
2008 Nov 22.0 & $WFPC2/PC$ & $F450W$ & 800 & 10877\\
2008 Nov 22.0 & $WFPC2/PC$ & $F675W$ & 360 & 10877\\
\\
2010 Apr 30.5 & $ACS/WFC1$ & $F658N$ & 1380 & 11675$^{2}$ \\
2010 Apr 30.6 & $ACS/WFC1$ & $F625W$ & 897 & 11675 \\
2010 Apr 30.6 & $ACS/WFC1$ & $F555W$ & 868 & 11675 \\
2010 Apr 30.6 & $ACS/WFC1$ & $F435W$ & 868 & 11675 \\
\hline\hline
\end{tabular}\\
$^{1}$ PI: W. Li\\
$^{2}$ PI: J. Maund\\
\end{table}

The site of SN~2006jc was observed at two separate epochs using the Hubble Space Telescope (HST) with the Wide Field Planetary Camera 2 (WFPC2) and the Advanced Camera for Surveys (ACS), and a log of these observations is presented in Table \ref{tab:obs}. \\ 
The WFPC2 observations from 2008 (or 776 days post-maximum) were retrieved from the Space Telescope Science Institute HST archive \footnote{https://archive.stsci.edu/hst/}, having been processed through the On-the-fly-recalibration pipeline. Photometry of the WFPC2 observations was conducted using the DOLPHOT package\footnote{http://americano.dolphinsim.com/dolphot/} \citep{dolphhstphot}, with the WFPC2 specific module.  The position of SN~2006jc fell on the Planetary Camera chip, which has a pixel scale of 0.05 arcsec. \\
The 2010 observations (1303 days post-maximum) were acquired using the $\mathrm{1k \times 1k}$ subarray of the ACS Wide Field Channel (WFC) 1. The observations, in each filter, were composed of four separate dithered exposures to aid in improving the sampling of the Point Spread Function (PSF).  The individual exposures were subject to ``bias striping noise", leading to obvious horizontal stripes across each image \citep{2010hstc.workE..54G}.  The horizontal noise features were almost completely removed using the ACS\_DESTRIPE\_PLUS package\footnote{http://www.stsci.edu/hst/acs/software/destripe/}, running in the PyRAF environment \footnote{PyRAF is a product of the Space Telescope Science Institute, which is operated by AURA for NASA.}, however at low levels some evidence of these stripes is just perceivable in the corrected images.  The observations were then processed and combined using the ASTRODRIZZLE package\footnote{http://drizzlepac.stsci.edu/}, which also corrects for the geometric distortion of the ACS WFC cameras.  We found that attempts to decrease the output pixel size to $\leq 0.03$ arcsec resulted in obvious aliasing patterns in the final combined images for each filter; and, therefore, we only drizzled the observations to a final pixel size of $0.05$ arcsec, matching the original ACS WFC pixel scale.  Photometry of the ACS observations was conducted using DOLPHOT with the ACS specific module. \\
For both the WFPC2 and ACS observations, photometric limits were determined using artificial star tests in which fake stars were inserted into the images and recovery was attempted using the same algorithm used for detecting and conducting photometry on real sources.  An artificial star was deemed to have been successfully detected if it was recovered within 1 pixel of the location at which it was inserted and its measured brightness was within $1\sigma$ of the input brightness.  The detection probability, as a function of magnitude, was parameterised as a cumulative normal distribution.\\
In order to determine the position of SN~2006jc on the HST observations, we  followed the procedure of \citet{2008MNRAS.389..141M}.  We measured the position of SN~2006jc in a deep 2008 Gemini Near Infrared Imager (NIRI) $K$-band image (with a precision of $0.75$ pixels).  The Gemini NIRI image was aligned with the 2010 ACS WFC $F555W$ observation, using $N=20$ stars, on which we were able to determine the SN location with a precision of $0.7$ pixels or 0.035 arcsec.  The position was further mapped (using $N= 17$ stars) to the 2008 WFPC2 $F555W$ observation, with a final uncertainty of $0.8$ pixels (or $0.04\,\mathrm{arcsec}$).

\section{Results and analysis}
\label{sec:res} 
\begin{figure}
\includegraphics[width=8cm]{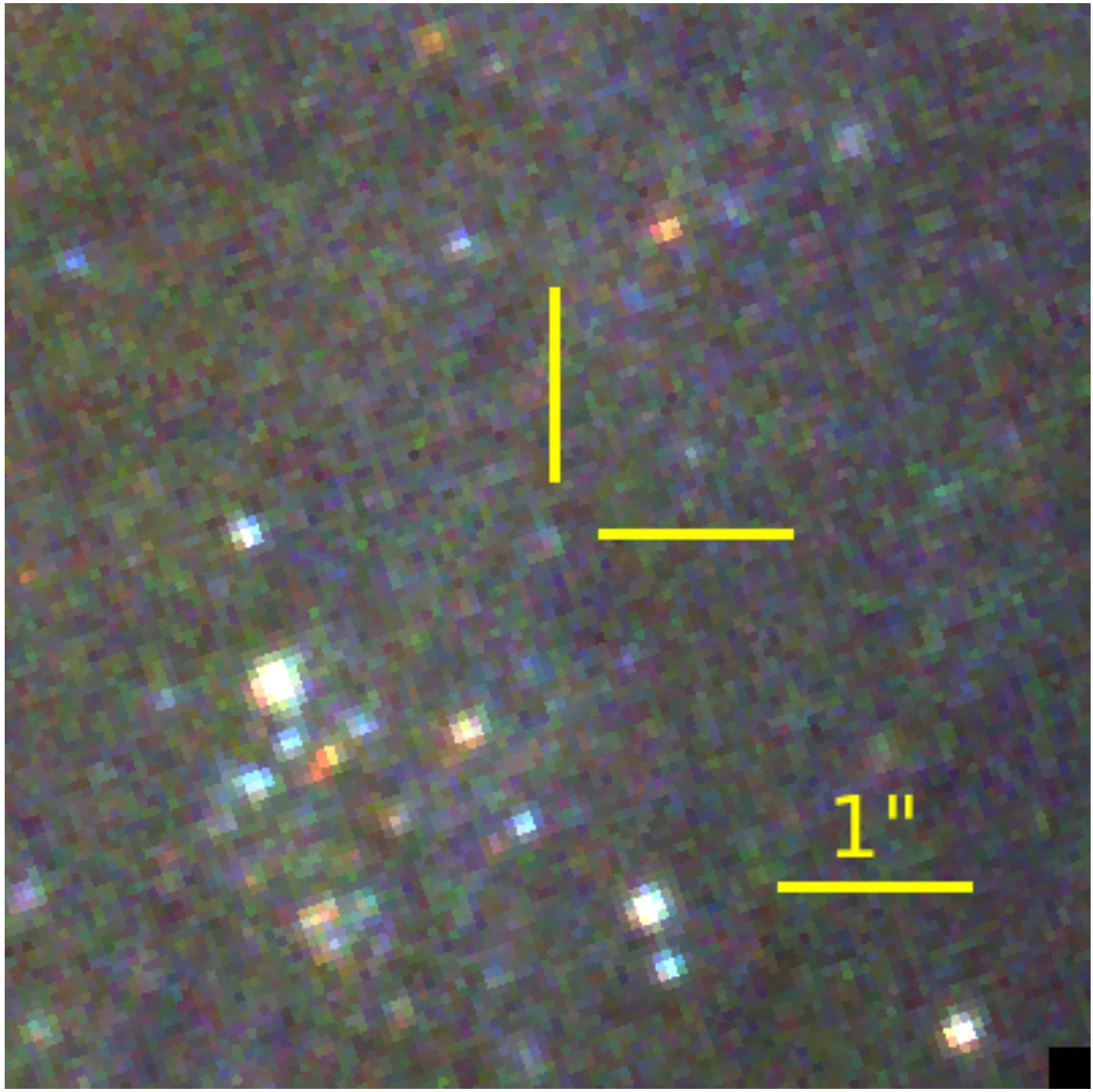}
\caption{HST ACS WFC observations of the site SN~2006jc acquired in 2010.  The three colour image is a composed of the $F435W$, $F555W$ and $F625W$ observations and the position of SN~2006jc is indicated by the cross-hairs. The image is oriented with North up and East to the left.}
\label{fig:res:image}
\end{figure}

\begin{figure}
\includegraphics[width=7.5cm, angle=270]{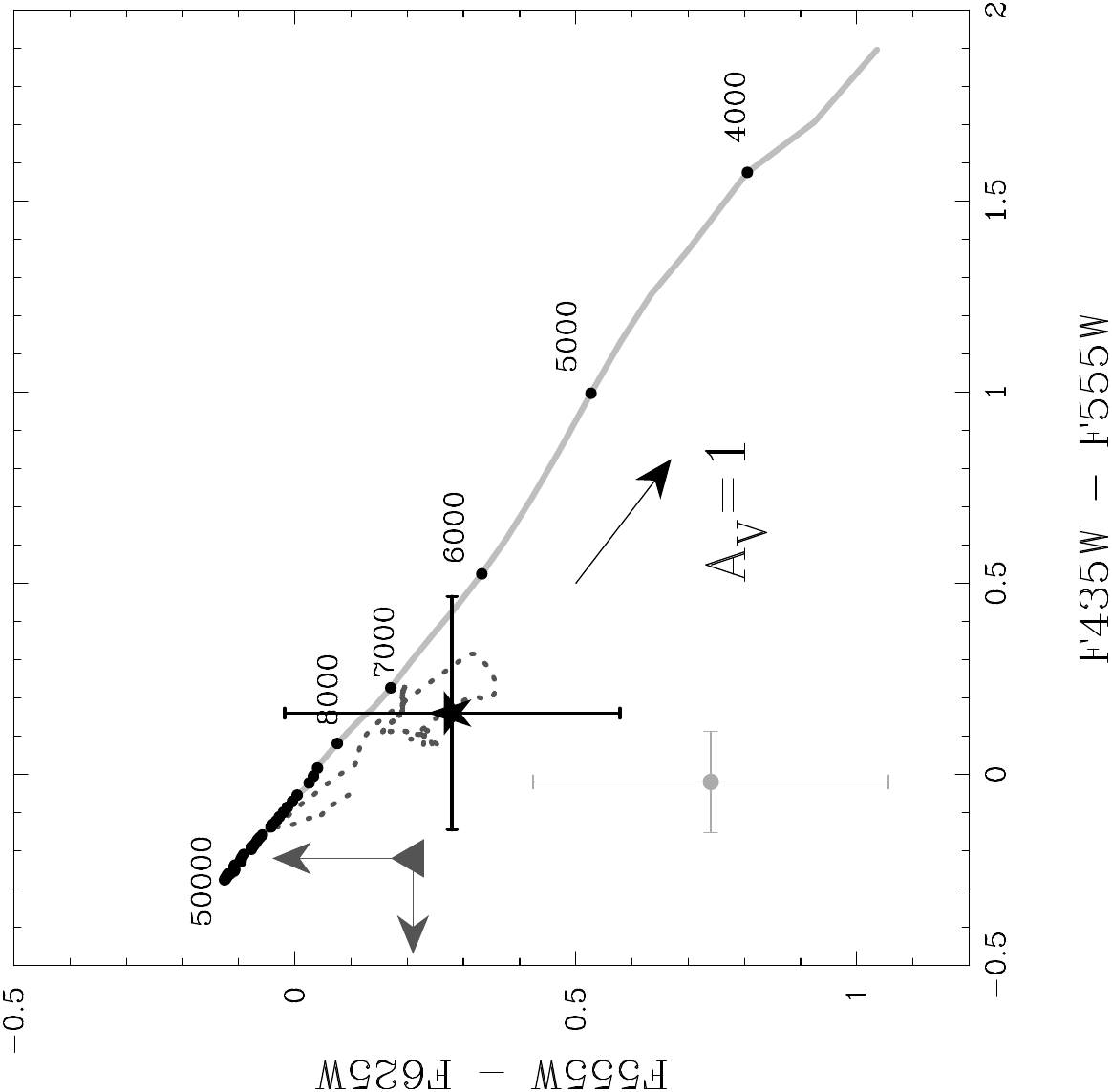}
\caption{A colour-colour diagram showing the photometry of the late-time source at the position of SN~2006jc (indicated by $\bigstar$), with respect to the two-colour sequences ($E(B-V) = 0$) for supergiants (solid grey)  in the temperature range $ 3500 \leq T \leq 50\,000\mathrm{K}$ and STARBURST99 cluster models (dotted dark grey line) for the age range $1 \leq t \leq 300\,\mathrm{Myr}$.  Points along the supergiant two-colour sequence delineate $1000\mathrm{K}$ intervals (as labeled).  Also shown, as the arrow, is the reddening vector corresponding to $A_{V} = 1$ mag.  The light grey point shows the last complete $BVR$ photometric measurement made by \citet{2008MNRAS.389..113P} at 109.7 days post-maximum.  The dark triangle ($\blacktriangle$) indicates the colour predicted for a light echo, derived from integrated colour of the observed light curve of SN~2006jc.}
\label{fig:res:2col}
\end{figure}

The site of SN~2006jc in the late-time HST ACS observations is shown in Figure \ref{fig:res:image}.  Although there is a large concentration of bright sources to the South East of the position of SN~2006jc, the SN position itself is remarkably isolated.  A source is recovered close to the transformed SN position (with an offset $\Delta r = 0.025$ arcsec or $0.5$ pixels) in the 2010 ACS observations with $m_{F435W} = 26.76\pm0.20$ ($S/N = 5.3$), $m_{F555W} = 26.60 \pm 0.23$ (4.7) and $m_{F625W} = 26.32\pm0.19$ (5.8) mags.  The values for sharpness and $\chi^{2}$ for the PSF-fitting photometry conducted by DOLPHOT suggested, for each of the three filters, that the late-time source was point-like and not extended.  The source is, however, not significantly detected ($S/N \geq 3$) in the $F658N$ observation,  which would include $\mathrm{H\alpha}$ emission at the redshift of the host galaxy, to a limit of $m_{F658N} = 24.25 \pm 0.05$ mags.  No source is recovered at the SN position in the 2008 WFPC2 observations, to limits of $m_{F450W}=24.88\pm 0.31$, $m_{F555W}=24.40\pm 0.58$, $m_{F675W} = 24.00\pm0.56$ and $m_{F814W} = 24.00\pm 0.87$ mags.  A comparison of the photometry of the late-time source with the colour-colour sequence for supergiants, derived using ATLAS9 synthetic spectra \citep[][with the parameters for supergiants suggested by \citealt{schmidtkaler}]{2004astro.ph..5087C},  and for stellar clusters using STARBURST99 \citep{1999ApJS..123....3L} is shown on Fig. \ref{fig:res:2col}.\\
Determining the exact nature of the late-time source at the position of SN~2006jc is complicated by the source's detection at only one epoch, and at very faint levels.  There are a number of possible scenarios for the origin of this flux, if it is actually associated with SN 2006jc: brightness arising from the SN itself at late-times; a light echo; a host cluster; an associated star; or a spatially coincident, but unassociated star.  \\

\begin{figure}
\includegraphics[width=8cm, angle=270]{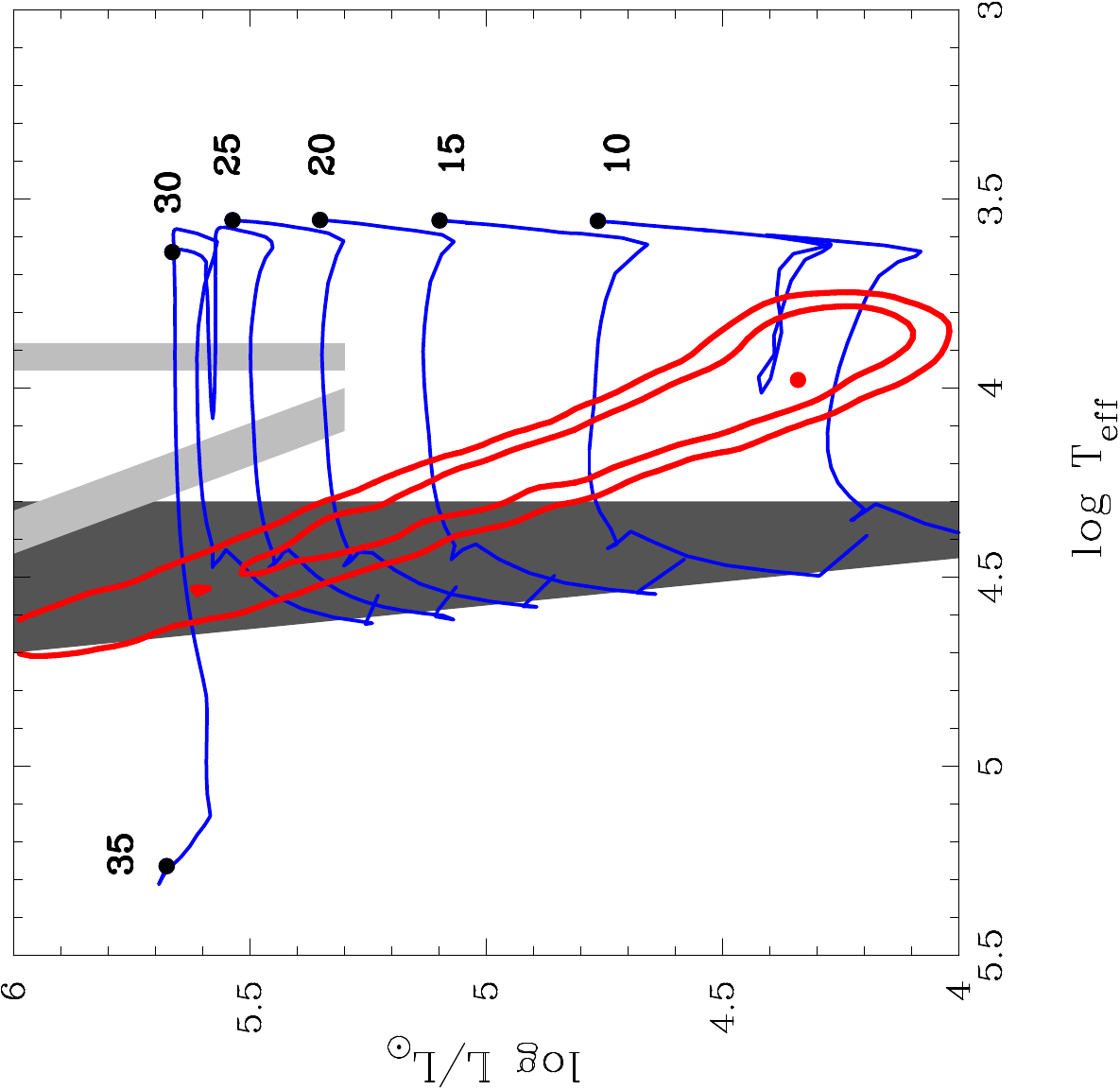}
\caption{Hertzsprung-Russell Diagram showing the position of the late-time source as derived from photometry of the 2010 ACS WFC observations, under the assumption that it is a stellar source.  The two contours presented contain 68\% and 95\% of the probability, while the red point indicates the most likely solution.  Overlaid are half-solar metallicity Cambridge STARS stellar evolution tracks.  The dark shaded region indicates the locus of binary companions, at the time of the death of the primary, produced by BPASS binary stellar evolution models.  The light shaded regions are the approximate locations for S Dor variables (at visual minimum and maximum) derived from \citet{2012ASSL..384..221V}.}
\label{fig:res:hrd}
\end{figure}

\citet{2007Natur.447..829P}, \citet{2008MNRAS.389..113P}, \citet{2007ApJ...657L.105F}, \citet{2008ApJ...674L..85I}, \citet{2009MNRAS.392..894A} and \citet{2014ApJS..213...19B} presented optical lightcurves of SN~2006jc covering up to 180d after maximum light, showing the SN 2006jc exhibited a precipitous decline in brightness ($\sim2\,\mathrm{mags}\,100\mathrm{d}^{-1}$ in the $V$-band).    If the light curve decay suddenly flattened at this time, for SN 2006jc to reach the brightness of the source observed in
2010 would require the decline rate to drop dramatically to $0.2\,\mathrm{mags}\,100\,\mathrm{d}^{-1}$ (or even lower if the flattening of the lightcurve occurred later).   The lack of a detection of SN~2006jc in the 2008 WFPC2 observations suggests that the rate of decline observed at 180d continued at much later epochs.  The last reported three-colour photometry of SN~2006jc in 2007 reported by \citet{2008MNRAS.389..113P}, in comparable bands to the 2010 observations,  also shows that SN 2006jc becoming redder in the $V-R$ colour  compared to the source recovered in 2010 (see Fig. \ref{fig:res:2col}).  Some caution, however, is required due to the large uncertainties associated with the later photometry of SN 2006jc and the photometry of the 2010 source.  Based on the red late-time colour of SN~2006jc and the severe implications for a very sudden flattening of the light curve after $\sim 180\mathrm{d}$ we consider it unlikely that the source recovered in the 2010 observations is SN 2006jc. 

The point-like nature of the late-time source (FWHM $= 2.3$ pix, corresponding to 15.5 pc at the distance of UGC 4904) is evidence against its origin in a light echo arising in an interstellar dust sheet, as it would require a very specific configuration of a sheet of scattering dust located immediately behind SN~2006jc.   Observations by  \citet{2008ApJ...680..568S}, \citet{2008MNRAS.389..141M} and \citet{2008ApJ...684..471D} indicated the presence of significant dust formation after $\sim 70\,\mathrm{days}$, with \citeauthor{2008MNRAS.389..141M} measuring $A_{V} = 2.8$ mags, arising from dust within $1000\,\mathrm{AU}$ of the point of the explosion, 230 days post-explosion. Based on observations at mid-infrared wavelengths \citeauthor{2008MNRAS.389..141M} also reported the presence of a pre-existing shell of dust extending to $\sim1\,\mathrm{pc}$ around the progenitor of SN~2006jc in which a light echo might arise and which, at its maximum extent, would remain unresolved in these late-time observations.  At the time of these observations, however, the light echo would arise {\it behind} SN~2006jc, requiring scattering through angles $>90^{\circ}$ which is inefficient \citep{2003ApJ...598.1017D}.

In addition, the contribution of an optical light echo is expected to produce a SED which is very close to that of the SN at early epochs.   To assess the likely colour of a light echo,  we considered the colour of SN 2006jc integrated over the entire light curve  reported by \citet{2007Natur.447..829P} and \citet{2008MNRAS.389..113P}.  We conservatively assume that the brightness from the time of explosion to the first photometric measurements, which were only acquired after maximum, is constant.  We find the total integrated colour, under these conditions, to be $B-V \sim -0.22$ and $V-I \sim 0.21$,  which is consistent with the earliest observed colours of $B-V = -0.15$ and $V-R=0.1$ mags \citep[see, e.g.,][]{2007Natur.447..829P} and bluer than the photometry of the 2010 source (see Fig. \ref{fig:res:2col}).  The evolution of the observed colour of SN~2006jc and other Type Ibn SNe \cite[e.g. SN 2010al;][]{2015MNRAS.449.1921P} suggest, however, the SN may have been even bluer prior to the first photometric measurements, implying that our integrated colour is in fact a red limit for the colour of a possible light echo.

As evident in Fig. \ref{fig:res:image}, SN~2006jc occurred in a relatively sparse area of UGC 4904.  In an area within 1 arcsec of the position of SN~2006jc, we find 10 sources recovered with $\mathrm{S/N \geq 5}$, including the source at the SN position.  Given the density of sources within the 1 arcsec area, we estimate the probability of finding one bright source inside the $1\sigma$ error radius on the position of SN~2006jc to be 0.01, implying that it is unlikely to find by chance an unrelated source to be spatially coincident with the SN position.\\
Overall, given the brightness, colours and spatial extent of the late-time source derived from the 2010 ACS observations, the simplest explanation is that this source is a star and is likely to be associated with SN~2006jc.  Future observations with HST will be required to confirm the continued presence of this source if it is indeed a star.    On the colour-colour plane (see Fig. \ref{fig:res:2col}), the position of the late-time source is coincident the supergiant two-colour sequence for limited reddening.    Assuming a Galactic $R_{V}=3.1$ \citealt{ccm89} reddening law and the total extinction measured towards SN 2006jc at early times by \citet{2007Natur.447..829P} of $A_{V} = 0.15$ mag, the corresponding probability density function for the position of the source on the Hertzsprung-Russell (HR) diagram, in conjunction with Cambridge STARS models \citep{eld04}, is shown on Fig. \ref{fig:res:hrd}.  If no prior is assumed for the reddening, the colours of the late-time source, assuming a stellar origin, limit the total reddening to $E(B-V)<0.6$ mag.  While the filter combination used for the 2010 ACS WFC observations can be used to place a constraint on the degree of reddening (and, therefore, extinction), it is not a sensitive diagnostic of stellar temperature at high temperatures.   The range of temperatures and luminosities that can be accommodated by the observed photometry primarily reflects the large range in possible bolometric corrections, rather than being due to uncertainties in reddening.  Despite the extent of possible locations for this source on the HR diagram, we note this star is outside of the regions associated with S Doradus variables in either quiescence or outburst \citep{2004ApJ...615..475S,2012ASSL..384..221V}. 

If the late-time source is an unresolved cluster, which might have hosted the progenitor of SN~2006jc, we note that absolute magnitude of the source, excluding considerations of extinction, is $M_{V} \sim -5.6$, which is 3 mags fainter than the suggested brightness at which unresolved clusters may be confused for individual stars  \citep{2005AA...443...79B}.    In comparison with STARBURST99 models \citep{1999ApJS..123....3L} the colours of the late-time source are consistent with an unresolved cluster for $E(B-V) \sim 0$ mags. If SN~2006jc occurred in a host cluster, any extinction that the cluster was subject to would also have affected the SN at early times, which was not observed.  In addition, there is insufficient time for any dust formed in SN 2006jc to be distributed on spatial scales similar to the typical sizes of clusters \citep[$\geq 1\,\mathrm{pc}$][]{2007A&A...469..925S} to extinguish a host cluster by the time of the 2010 observations.

\section{Discussion \& Conclusions}
\label{sec:disc}

\begin{figure}
\includegraphics[width=6.1cm, angle=270]{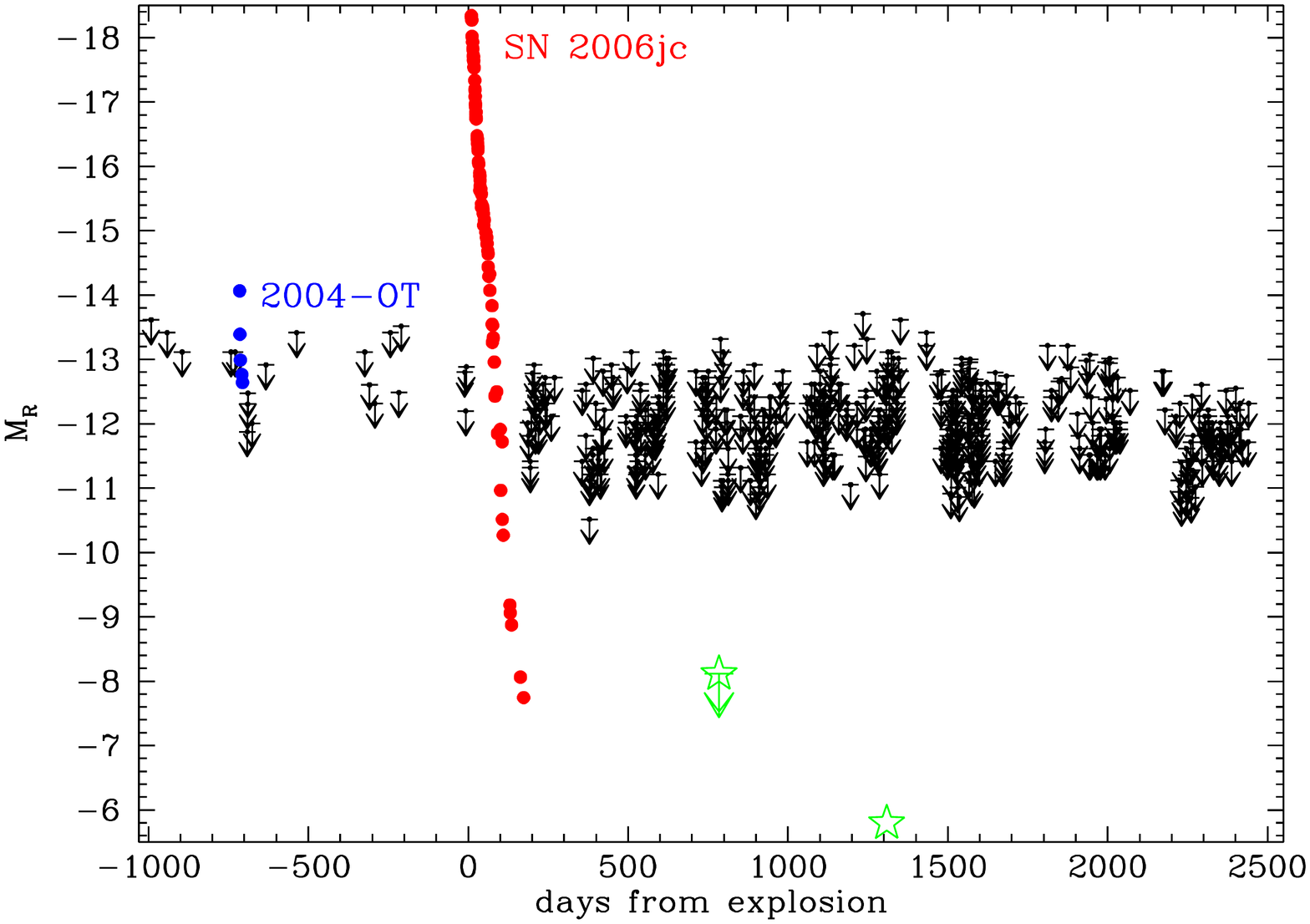}
\caption{Longterm monitoring of the position of SN~2006jc in UGC 4904, with dates given with respect to the explosion date of the SN.  Black arrows indicate upper limits on the brightness at the SN position derived from regular monitoring of amateur astronomers and from PTF images \citep[second data release][]{2009PASP..121.1395L} obtained through the Infrared Processing and Analysis Center interface ({\it http://www.ipac.caltech.edu/}); red and blue points correspond to the light curves of SN 2006jc and the 2004 optical transient precursor, respectively \citep{2007Natur.447..829P}; and green stars indicate limits and detections derived from late-time HST observations, as presented here.  The light curve of SN~2006jc plotted here is from data published by \citet{2007Natur.447..829P}, \citet{2007ApJ...657L.105F},\citet{2008MNRAS.389..113P}, \citet{2009MNRAS.392..894A} and \citet{2014ApJS..213...19B}.}
\label{fig:disc:monit}
\end{figure}

Based on the properties of the late-time source we cautiously conclude that the most likely explanation is that it is a star and the companion to the progenitor that exploded as SN~2006jc. The exact nature of the star is difficult to constrain given the available late-time HST photometry.  The fact that the locus of the star does not correspond exactly with the two locations in which S Doradus variables are found on the HR diagram would argue against this star being an LBV, but rather a normal supergiant.  This conclusion is further supported by the absence of strong $\mathrm{H\alpha}$ emission at late-times as might be expected of classical LBVs; however we note that UGC2773-2009OT and LMC R71 also do not show strong $\mathrm{H\alpha}$ at late-times afters outbursts and that the limited depth of the late-time $F658N$ observation might not probe weak $
\mathrm{H\alpha}$ if it were present.\\
An important question raised by the possible binary nature of the progenitor system of SN~2006jc is which component was responsible for the 2004 precursor outburst.  The late-time HST photometry, at the lowest luminosity limit, is consistent with a $\sim10M_{\odot}$ A-F supergiant (see Fig. \ref{fig:res:hrd}), which would make it comparable to properties of the progenitors identified for faint outburst events such as SN 2008S and NGC 300-OT \citep{2008ApJ...681L...9P,2009ApJ...695L.154B,2009ApJ...699.1850B,2009MNRAS.398.1041B,2009ApJ...705.1364T,2012ApJ...750...77S}, but still fainter and less massive than the A-to-F-type hypergiant progenitors of more classical SN impostors \citep[e.g.][]{2010AJ....139.1451S,2015AA...581L...4K,2016arXiv160404628T}.  In direct contrast to the former two events, however, there is no evidence that the late-time source at the position of SN~2006jc is enshrouded in significant quantities of dust.  Furthermore, apart from the precursor and SN 2006jc, at the SN position, there is no evidence for any significant outburst events after the explosion of the SN.  Long term monitoring of the site of SN~2006jc, before and after the  2004 outburst and the 2006 SN explosion, is shown in Fig. \ref{fig:disc:monit}.  The limited history of outbursts at the location of SN~2006jc would suggest, therefore, that any significant variability ended with the SN.\\

It is interesting to note that the location of SN~2006jc is offset (by $\sim 2\,\mathrm{arcsec}$ or a projected distance of $\sim 270\,\mathrm{pc}$) from the nearest obvious sites containing young massive stars (see Fig. \ref{fig:res:image}).  Similar conditions have been observed for the hydrogen-rich interacting Type IIn SNe and some LBVs \citep{2015MNRAS.447..598S}; although \citet{2016arXiv160301278H}  note that classical LBVs are found in close proximity to late-type O stars while non-classical LBVs are found in relative isolation  (this matter was further debated in the exchange between \citet{2016MNRAS.461.3353S} and \citet{2016arXiv160802007D}).  The absence of significant $\mathrm{H\alpha}$ emission in the environment of SN~2006jc is also consistent with the observations of \citet{2008MNRAS.390.1527A} for the positions of interacting SNe with respect to H\,{\sc ii} regions.  This could support the interpretation of the properties of the late-time source as a lower mass star ($\sim 10M_{\odot}$), that is not associated with recent massive star formation.  The companion would then have to have been sufficiently close to the progenitor to interact with it to induce the outburst; however, the absence of strong H features from the subsequent SN spectra would require the companion to be sufficiently distant at the time of explosion (e.g. $\sim 0.01M_{\odot}$ of H in Type IIb SNe;  although we note that \citet{2007Natur.447..829P}, \citet{2007ApJ...657L.105F} and \citet{2008ApJ...680..568S} report the emergence of weak $H\alpha$ emission at $\sim$51 days which could be related to the putative companion star).

 We compared the position of the late-time source on the HR diagram with the locus of binary companions, at the time of the explosion of the primary, predicted by Binary Population and Spectral Synthesis (BPASS\footnote{http://bpass.auckland.ac.nz/}) models \citep{2009MNRAS.400.1019E}, as shown on Figure \ref{fig:res:hrd} \citep[see also][]{2016MNRAS.461L.117E}.   The locus for the companions, generally following the main sequence and its turn-off, intersects with the contours for the late-time source at higher masses  ($\gtrsim 20M_{\odot}$).  If the companion was indeed massive, this would favour a high mass WR progenitor.  We caution, however, that this family of models do not predict the progenitor and binary companion for SN~1993J, given the observed constraints on the binary progenitor system determined by \citet{maund93j}.

\citet{2014MNRAS.445.2492M} present a model whereby an outburst can be triggered by the interaction of a main sequence companion star with the extended envelope of an evolved primary,  with eccentric orbits making the outbursts more extreme.  \citet{2011MNRAS.415.2020S} proposes a similar model, however it requires a collision between the two stars to produce an outburst; which cannot be the case for SN~2006jc due to the persistence of the source at the SN position after the explosion.  An interesting consequence of repeated periastron passages noted by \citet{2011MNRAS.415.2020S} is the removal of H envelope of the primary and the increasing eccentricity of the orbit, with \citeauthor{2011MNRAS.415.2020S} specifically noting the case of WR 140.  The dependence of the mass loss history of the primary on the orbital properties of the binary system might also explain the evidence for episodic mass loss in the form of shells of material discernible in early-time spectra of SN 2006jc \citep{2007Natur.447..829P}.  Given the importance of binary interactions in the evolution of massive stars \citep{2012Sci...337..444S} previous encounters with the companion may also have responsible for the H-poor nature of the progenitor and not just the single observed pre-explosion outburst. 

An additional possible progenitor scenario for SN 2006jc is suggested by the successful identification \citep{2013ApJ...775L...7C} and confirmation, through disappearance, of the low mass progenitor of the Type Ib SN iPTF 13bvn \citep{2016MNRAS.461L.117E,2016ApJ...825L..22F}.  In this scenario the progenitor of SN~2006jc need not have been a massive star capable of evolving into a WR star or an LBV, but was rather a lower mass star ($M_{ZAMS} \sim 8M_{\odot}$) that was stripped of its H envelope and that then evolved as a helium giant.  The luminosities of such stars would be insufficient to drive strong winds, comparable to higher mass WR stars, but would instead have extended He-rich envelopes \citep{2016MNRAS.459.1505M}.  Due to the structural differences between helium giants and massive WR stars, the interaction between a low mass helium giant and a binary companion could be different to that expected for a binary system containing a WR star, and might be more amenable to the production of pre-explosion outbursts and the laying down of a dense CSM.  Detailed calculations of such a binary configuration are, however, beyond the scope of this paper.   As noted above, however, the predicted locus for the low mass binary companion to the progenitor of iPTF 13bvn presented by \citet{2016MNRAS.461L.117E} lies at higher temperatures than we have derived for the source recovered in the 2010 observations. 

Ultimately, confirmation with future HST observations of the SN location to required to test if the source is still present or if it has disappeared and to identify which of the above scenarios is responsible for the late-time brightness of SN~2006jc.

\section*{Acknowledgments} 
The research of JRM is supported through a Royal Society University Research Fellowship.  AP is partially supported by the PRIN-INAF 2014 with the project ``Transient Universe: unveiling new types of stellar explosions with PESSTO''.  We thank Noam Soker for very useful discussions during the preparation of the manuscript.  We thank the team members of the Italian Supernovae Search Project (ISSP: http://italiansupernovae.org/), in particular Fabio Briganti and Fabio Martinelli  of the Astronomical Centre, Lajatico, Pisa (http://www.astronomicalcentre.org/) and Fabrizio Ciabattari of the Astronomical Observatory of Monte Agliale, Borgo a Mozzano, Lucca (http://www.oama.it/) for sharing their images of UGC 4904 with us. We are also grateful to T. Yusa for his help in collecting data from amateur astronomers.   This research has made use of the NASA/ IPAC Infrared Science Archive, which is operated by the Jet Propulsion Laboratory, California Institute of Technology, under contract with the National Aeronautics and Space Administration.\\
\facilities{HST(ACS)}

\bibliographystyle{aasjournal}

\begin{thebibliography}{}
\expandafter\ifx\csname natexlab\endcsname\relax\def\natexlab#1{#1}\fi

\bibitem[{{Anderson} \& {James}(2008)}]{2008MNRAS.390.1527A}
{Anderson}, J.~P., \& {James}, P.~A. 2008, \mnras, 390, 1527

\bibitem[{{Anupama} {et~al.}(2009){Anupama}, {Sahu}, {Gurugubelli}, {Prabhu},
  {Tominaga}, {Tanaka}, \& {Nomoto}}]{2009MNRAS.392..894A}
{Anupama}, G.~C., {Sahu}, D.~K., {Gurugubelli}, U.~K., {et~al.} 2009, \mnras,
  392, 894

\bibitem[{{Bastian} {et~al.}(2005){Bastian}, {Gieles}, {Efremov}, \&
  {Lamers}}]{2005AA...443...79B}
{Bastian}, N., {Gieles}, M., {Efremov}, Y.~N., \& {Lamers}, H.~J.~G.~L.~M.
  2005, \aap, 443, 79

\bibitem[{{Berger} {et~al.}(2009){Berger}, {Soderberg}, {Chevalier},
  {Fransson}, {Foley}, {Leonard}, {Debes}, {Diamond-Stanic}, {Dupree}, {Ivans},
  {Simmerer}, {Thompson}, \& {Tremonti}}]{2009ApJ...699.1850B}
{Berger}, E., {Soderberg}, A.~M., {Chevalier}, R.~A., {et~al.} 2009, \apj, 699,
  1850

\bibitem[{{Bianco} {et~al.}(2014){Bianco}, {Modjaz}, {Hicken}, {Friedman},
  {Kirshner}, {Bloom}, {Challis}, {Marion}, {Wood-Vasey}, \&
  {Rest}}]{2014ApJS..213...19B}
{Bianco}, F.~B., {Modjaz}, M., {Hicken}, M., {et~al.} 2014, \apjs, 213, 19

\bibitem[{{Bond} {et~al.}(2009){Bond}, {Bedin}, {Bonanos}, {Humphreys},
  {Monard}, {Prieto}, \& {Walter}}]{2009ApJ...695L.154B}
{Bond}, H.~E., {Bedin}, L.~R., {Bonanos}, A.~Z., {et~al.} 2009, \apjl, 695,
  L154

\bibitem[{{Botticella} {et~al.}(2009){Botticella}, {Pastorello}, {Smartt},
  {Meikle}, {Benetti}, {Kotak}, {Cappellaro}, {Crockett}, {Mattila}, {Sereno},
  {Patat}, {Tsvetkov}, {van Loon}, {Abraham}, {Agnoletto}, {Arbour}, {Benn},
  {di Rico}, {Elias-Rosa}, {Gorshanov}, {Harutyunyan}, {Hunter}, {Lorenzi},
  {Keenan}, {Maguire}, {Mendez}, {Mobberley}, {Navasardyan}, {Ries},
  {Stanishev}, {Taubenberger}, {Trundle}, {Turatto}, \&
  {Volkov}}]{2009MNRAS.398.1041B}
{Botticella}, M.~T., {Pastorello}, A., {Smartt}, S.~J., {et~al.} 2009, \mnras,
  398, 1041

\bibitem[{{Cao} {et~al.}(2013){Cao}, {Kasliwal}, {Arcavi}, {Horesh}, {Hancock},
  {Valenti}, {Cenko}, {Kulkarni}, {Gal-Yam}, {Gorbikov}, {Ofek}, {Sand},
  {Yaron}, {Graham}, {Silverman}, {Wheeler}, {Marion}, {Walker}, {Mazzali},
  {Howell}, {Li}, {Kong}, {Bloom}, {Nugent}, {Surace}, {Masci}, {Carpenter},
  {Degenaar}, \& {Gelino}}]{2013ApJ...775L...7C}
{Cao}, Y., {Kasliwal}, M.~M., {Arcavi}, I., {et~al.} 2013, \apjl, 775, L7

\bibitem[{{Cardelli} {et~al.}(1989){Cardelli}, {Clayton}, \& {Mathis}}]{ccm89}
{Cardelli}, J.~A., {Clayton}, G.~C., \& {Mathis}, J.~S. 1989, \apj, 345, 245

\bibitem[{{Castelli} \& {Kurucz}(2004)}]{2004astro.ph..5087C}
{Castelli}, F., \& {Kurucz}, R.~L. 2004, ArXiv Astrophysics e-prints,
  arXiv:astro-ph/0405087

\bibitem[{{Davidson} {et~al.}(2016){Davidson}, {Humphreys}, \&
  {Weis}}]{2016arXiv160802007D}
{Davidson}, K., {Humphreys}, R.~M., \& {Weis}, K. 2016, ArXiv e-prints,
  arXiv:1608.02007

\bibitem[{{Di Carlo} {et~al.}(2008){Di Carlo}, {Corsi}, {Arkharov}, {Massi},
  {Larionov}, {Efimova}, {Dolci}, {Napoleone}, \& {Di
  Paola}}]{2008ApJ...684..471D}
{Di Carlo}, E., {Corsi}, C., {Arkharov}, A.~A., {et~al.} 2008, \apj, 684, 471

\bibitem[{{Dolphin}(2000)}]{dolphhstphot}
{Dolphin}, A.~E. 2000, \pasp, 112, 1383

\bibitem[{{Draine}(2003)}]{2003ApJ...598.1017D}
{Draine}, B.~T. 2003, \apj, 598, 1017

\bibitem[{{Eldridge} \& {Maund}(2016)}]{2016MNRAS.461L.117E}
{Eldridge}, J.~J., \& {Maund}, J.~R. 2016, \mnras, 461, L117

\bibitem[{{Eldridge} \& {Stanway}(2009)}]{2009MNRAS.400.1019E}
{Eldridge}, J.~J., \& {Stanway}, E.~R. 2009, \mnras, 400, 1019

\bibitem[{{Eldridge} \& {Tout}(2004)}]{eld04}
{Eldridge}, J.~J., \& {Tout}, C.~A. 2004, \mnras, 353, 87

\bibitem[{{Folatelli} {et~al.}(2016){Folatelli}, {Van Dyk}, {Kuncarayakti},
  {Maeda}, {Bersten}, {Nomoto}, {Pignata}, {Hamuy}, {Quimby}, {Zheng},
  {Filippenko}, {Clubb}, {Smith}, {Elias-Rosa}, {Foley}, \&
  {Miller}}]{2016ApJ...825L..22F}
{Folatelli}, G., {Van Dyk}, S.~D., {Kuncarayakti}, H., {et~al.} 2016, \apjl,
  825, L22

\bibitem[{{Foley} {et~al.}(2007){Foley}, {Smith}, {Ganeshalingam}, {Li},
  {Chornock}, \& {Filippenko}}]{2007ApJ...657L.105F}
{Foley}, R.~J., {Smith}, N., {Ganeshalingam}, M., {et~al.} 2007, \apjl, 657,
  L105

\bibitem[{{Gal-Yam} \& {Leonard}(2009)}]{galyam05gl}
{Gal-Yam}, A., \& {Leonard}, D.~C. 2009, \nat, 458, 865

\bibitem[{{Grogin} {et~al.}(2010){Grogin}, {Lim}, {Maybhate}, {Hook}, \&
  {Loose}}]{2010hstc.workE..54G}
{Grogin}, N.~A., {Lim}, P.~L., {Maybhate}, A., {Hook}, R.~N., \& {Loose}, M.
  2010, in Hubble after SM4. Preparing JWST, 54

\bibitem[{{Hosseinzadeh} {et~al.}(2016){Hosseinzadeh}, {Arcavi}, {Howell},
  {McCully}, \& {Valenti}}]{2016AAS...22712003H}
{Hosseinzadeh}, G., {Arcavi}, I., {Howell}, D.~A., {McCully}, C., \& {Valenti},
  S. 2016, in American Astronomical Society Meeting Abstracts, Vol. 227,
  American Astronomical Society Meeting Abstracts, 120.03

\bibitem[{{Humphreys} {et~al.}(2016){Humphreys}, {Weis}, {Davidson}, \&
  {Gordon}}]{2016arXiv160301278H}
{Humphreys}, R.~M., {Weis}, K., {Davidson}, K., \& {Gordon}, M.~S. 2016, ArXiv
  e-prints, arXiv:1603.01278

\bibitem[{{Immler} {et~al.}(2008){Immler}, {Modjaz}, {Landsman}, {Bufano},
  {Brown}, {Milne}, {Dessart}, {Holland}, {Koss}, {Pooley}, {Kirshner},
  {Filippenko}, {Panagia}, {Chevalier}, {Mazzali}, {Gehrels}, {Petre},
  {Burrows}, {Nousek}, {Roming}, {Pian}, {Soderberg}, \&
  {Greiner}}]{2008ApJ...674L..85I}
{Immler}, S., {Modjaz}, M., {Landsman}, W., {et~al.} 2008, \apjl, 674, L85

\bibitem[{{Kankare} {et~al.}(2015){Kankare}, {Kotak}, {Pastorello}, {Fraser},
  {Mattila}, {Smartt}, {Bruce}, {Chambers}, {Elias-Rosa}, {Flewelling},
  {Fremling}, {Harmanen}, {Huber}, {Jerkstrand}, {Kangas}, {Kuncarayakti},
  {Magee}, {Magnier}, {Polshaw}, {Smith}, {Sollerman}, \&
  {Tomasella}}]{2015AA...581L...4K}
{Kankare}, E., {Kotak}, R., {Pastorello}, A., {et~al.} 2015, \aap, 581, L4

\bibitem[{{Law} {et~al.}(2009){Law}, {Kulkarni}, {Dekany}, {Ofek}, {Quimby},
  {Nugent}, {Surace}, {Grillmair}, {Bloom}, {Kasliwal}, {Bildsten}, {Brown},
  {Cenko}, {Ciardi}, {Croner}, {Djorgovski}, {van Eyken}, {Filippenko}, {Fox},
  {Gal-Yam}, {Hale}, {Hamam}, {Helou}, {Henning}, {Howell}, {Jacobsen},
  {Laher}, {Mattingly}, {McKenna}, {Pickles}, {Poznanski}, {Rahmer}, {Rau},
  {Rosing}, {Shara}, {Smith}, {Starr}, {Sullivan}, {Velur}, {Walters}, \&
  {Zolkower}}]{2009PASP..121.1395L}
{Law}, N.~M., {Kulkarni}, S.~R., {Dekany}, R.~G., {et~al.} 2009, \pasp, 121,
  1395

\bibitem[{{Leitherer} {et~al.}(1999){Leitherer}, {Schaerer}, {Goldader},
  {Gonz{\'a}lez Delgado}, {Robert}, {Kune}, {de Mello}, {Devost}, \&
  {Heckman}}]{1999ApJS..123....3L}
{Leitherer}, C., {Schaerer}, D., {Goldader}, J.~D., {et~al.} 1999, \apjs, 123,
  3

\bibitem[{{Mattila} {et~al.}(2008){Mattila}, {Meikle}, {Lundqvist},
  {Pastorello}, {Kotak}, {Eldridge}, {Smartt}, {Adamson}, {Gerardy}, {Rizzi},
  {Stephens}, \& {van Dyk}}]{2008MNRAS.389..141M}
{Mattila}, S., {Meikle}, W.~P.~S., {Lundqvist}, P., {et~al.} 2008, \mnras, 389,
  141

\bibitem[{{Maund} {et~al.}(2004){Maund}, {Smartt}, {Kudritzki},
  {Podsiadlowski}, \& {Gilmore}}]{maund93j}
{Maund}, J.~R., {Smartt}, S.~J., {Kudritzki}, R.~P., {Podsiadlowski}, P., \&
  {Gilmore}, G.~F. 2004, \nat, 427, 129

\bibitem[{{McClelland} \& {Eldridge}(2016)}]{2016MNRAS.459.1505M}
{McClelland}, L.~A.~S., \& {Eldridge}, J.~J. 2016, \mnras, 459, 1505

\bibitem[{{Mcley} \& {Soker}(2014)}]{2014MNRAS.445.2492M}
{Mcley}, L., \& {Soker}, N. 2014, \mnras, 445, 2492

\bibitem[{{Nakano} {et~al.}(2006){Nakano}, {Itagaki}, {Puckett}, \&
  {Gorelli}}]{2006CBET..666....1N}
{Nakano}, S., {Itagaki}, K., {Puckett}, T., \& {Gorelli}, R. 2006, Central
  Bureau Electronic Telegrams, 666

\bibitem[{{Pastorello} {et~al.}(2007){Pastorello}, {Smartt}, {Mattila},
  {Eldridge}, {Young}, {Itagaki}, {Yamaoka}, {Navasardyan}, {Valenti}, {Patat},
  {Agnoletto}, {Augusteijn}, {Benetti}, {Cappellaro}, {Boles}, {Bonnet-Bidaud},
  {Botticella}, {Bufano}, {Cao}, {Deng}, {Dennefeld}, {Elias-Rosa},
  {Harutyunyan}, {Keenan}, {Iijima}, {Lorenzi}, {Mazzali}, {Meng}, {Nakano},
  {Nielsen}, {Smoker}, {Stanishev}, {Turatto}, {Xu}, \&
  {Zampieri}}]{2007Natur.447..829P}
{Pastorello}, A., {Smartt}, S.~J., {Mattila}, S., {et~al.} 2007, \nat, 447, 829

\bibitem[{{Pastorello} {et~al.}(2008){Pastorello}, {Mattila}, {Zampieri},
  {Della Valle}, {Smartt}, {Valenti}, {Agnoletto}, {Benetti}, {Benn}, {Branch},
  {Cappellaro}, {Dennefeld}, {Eldridge}, {Gal-Yam}, {Harutyunyan}, {Hunter},
  {Kjeldsen}, {Lipkin}, {Mazzali}, {Milne}, {Navasardyan}, {Ofek}, {Pian},
  {Shemmer}, {Spiro}, {Stathakis}, {Taubenberger}, {Turatto}, \&
  {Yamaoka}}]{2008MNRAS.389..113P}
{Pastorello}, A., {Mattila}, S., {Zampieri}, L., {et~al.} 2008, \mnras, 389,
  113

\bibitem[{{Pastorello} {et~al.}(2015){Pastorello}, {Benetti}, {Brown},
  {Tsvetkov}, {Inserra}, {Taubenberger}, {Tomasella}, {Fraser}, {Rich},
  {Botticella}, {Bufano}, {Cappellaro}, {Ergon}, {Gorbovskoy}, {Harutyunyan},
  {Huang}, {Kotak}, {Lipunov}, {Magill}, {Miluzio}, {Morrell}, {Ochner},
  {Smartt}, {Sollerman}, {Spiro}, {Stritzinger}, {Turatto}, {Valenti}, {Wang},
  {Wright}, {Yurkov}, {Zampieri}, \& {Zhang}}]{2015MNRAS.449.1921P}
{Pastorello}, A., {Benetti}, S., {Brown}, P.~J., {et~al.} 2015, \mnras, 449,
  1921

\bibitem[{{Pastorello} {et~al.}(2016){Pastorello}, {Wang}, {Ciabattari},
  {Bersier}, {Mazzali}, {Gao}, {Xu}, {Zhang}, {Tokuoka}, {Benetti},
  {Cappellaro}, {Elias-Rosa}, {Harutyunyan}, {Huang}, {Miluzio}, {Mo},
  {Ochner}, {Tartaglia}, {Terreran}, {Tomasella}, \&
  {Turatto}}]{2016MNRAS.456..853P}
{Pastorello}, A., {Wang}, X.-F., {Ciabattari}, F., {et~al.} 2016, \mnras, 456,
  853

\bibitem[{{Prieto} {et~al.}(2008){Prieto}, {Kistler}, {Thompson}, {Y{\"u}ksel},
  {Kochanek}, {Stanek}, {Beacom}, {Martini}, {Pasquali}, \&
  {Bechtold}}]{2008ApJ...681L...9P}
{Prieto}, J.~L., {Kistler}, M.~D., {Thompson}, T.~A., {et~al.} 2008, \apjl,
  681, L9

\bibitem[{{Sana} {et~al.}(2012){Sana}, {de Mink}, {de Koter}, {Langer},
  {Evans}, {Gieles}, {Gosset}, {Izzard}, {Le Bouquin}, \&
  {Schneider}}]{2012Sci...337..444S}
{Sana}, H., {de Mink}, S.~E., {de Koter}, A., {et~al.} 2012, Science, 337, 444

\bibitem[{{Sanders} {et~al.}(2013){Sanders}, {Soderberg}, {Foley}, {Chornock},
  {Milisavljevic}, {Margutti}, {Drout}, {Moe}, {Berger}{et~al.}}]{2013ApJ...769...39S}
{Sanders}, N.~E., {Soderberg}, A.~M., {Foley}, R.~J., {et~al.} 2013, \apj, 769,
  39

\bibitem[{{Scheepmaker} {et~al.}(2007){Scheepmaker}, {Haas}, {Gieles},
  {Bastian}, {Larsen}, \& {Lamers}}]{2007A&A...469..925S}
{Scheepmaker}, R.~A., {Haas}, M.~R., {Gieles}, M., {et~al.} 2007, \aap, 469,
  925

\bibitem[{Schmidt-Kaler(1982)}]{schmidtkaler}
Schmidt-Kaler, T. 1982, in Landolt-B{\"o}rnstein - Group VI Astronomy and
  Astrophysics, Vol.~2b, Stars and Star Clusters, ed. K.~Schaifers \& H.~Voigt
  (Springer Berlin Heidelberg), 14--24

\bibitem[{{Smith}(2011)}]{2011MNRAS.415.2020S}
{Smith}, N. 2011, \mnras, 415, 2020

\bibitem[{{Smith}(2014)}]{2014ARA&A..52..487S}
---. 2014, \araa, 52, 487

\bibitem[{{Smith}(2016)}]{2016MNRAS.461.3353S}
---. 2016, \mnras, 461, 3353

\bibitem[{{Smith} {et~al.}(2008){Smith}, {Foley}, \&
  {Filippenko}}]{2008ApJ...680..568S}
{Smith}, N., {Foley}, R.~J., \& {Filippenko}, A.~V. 2008, \apj, 680, 568

\bibitem[{{Smith} \& {Tombleson}(2015)}]{2015MNRAS.447..598S}
{Smith}, N., \& {Tombleson}, R. 2015, \mnras, 447, 598

\bibitem[{{Smith} {et~al.}(2004){Smith}, {Vink}, \& {de
  Koter}}]{2004ApJ...615..475S}
{Smith}, N., {Vink}, J.~S., \& {de Koter}, A. 2004, \apj, 615, 475

\bibitem[{{Smith} {et~al.}(2010){Smith}, {Miller}, {Li}, {Filippenko},
  {Silverman}, {Howard}, {Nugent}, {Marcy}, {Bloom}, {Ghez}, {Lu}, {Yelda},
  {Bernstein}, \& {Colucci}}]{2010AJ....139.1451S}
{Smith}, N., {Miller}, A., {Li}, W., {et~al.} 2010, \aj, 139, 1451

\bibitem[{{Szczygie{\l}} {et~al.}(2012){Szczygie{\l}}, {Prieto}, {Kochanek},
  {Stanek}, {Thompson}, {Beacom}, {Garnavich}, \&
  {Woodward}}]{2012ApJ...750...77S}
{Szczygie{\l}}, D.~M., {Prieto}, J.~L., {Kochanek}, C.~S., {et~al.} 2012, \apj,
  750, 77

\bibitem[{{Taddia} {et~al.}(2015){Taddia}, {Sollerman}, {Fremling},
  {Pastorello}, {Leloudas}, {Fransson}, {Nyholm}, {Stritzinger}, {Ergon},
  {Roy}, \& {Migotto}}]{2015AA...580A.131T}
{Taddia}, F., {Sollerman}, J., {Fremling}, C., {et~al.} 2015, \aap, 580, A131

\bibitem[{{Tartaglia} {et~al.}(2016){Tartaglia}, {Elias-Rosa}, {Pastorello},
  {Benetti}, {Taubenberger}, {Cappellaro}, {Cortini}, {Granata}, {Ishida},
  {Morales-Garoffolo}, {Noebauer}, {Ochner}, {Tomasella}, \&
  {Zaggia}}]{2016arXiv160404628T}
{Tartaglia}, L., {Elias-Rosa}, N., {Pastorello}, A., {et~al.} 2016, ArXiv
  e-prints, arXiv:1604.04628

\bibitem[{{Thompson} {et~al.}(2009){Thompson}, {Prieto}, {Stanek}, {Kistler},
  {Beacom}, \& {Kochanek}}]{2009ApJ...705.1364T}
{Thompson}, T.~A., {Prieto}, J.~L., {Stanek}, K.~Z., {et~al.} 2009, \apj, 705,
  1364

\bibitem[{{Tominaga} {et~al.}(2008){Tominaga}, {Limongi}, {Suzuki}, {Tanaka},
  {Nomoto}, {Maeda}, {Chieffi}, {Tornambe}, {Minezaki}, {Yoshii}, {Sakon},
  {Wada}, {Ohyama}, {Tanab{\'e}}, {Kaneda}, {Onaka}, {Nozawa}, {Kozasa},
  {Kawabata}, {Anupama}, {Sahu}, {Gurugubelli}, {Prabhu}, \&
  {Deng}}]{2008ApJ...687.1208T}
{Tominaga}, N., {Limongi}, M., {Suzuki}, T., {et~al.} 2008, \apj, 687, 1208

\bibitem[{{Vink}(2012)}]{2012ASSL..384..221V}
{Vink}, J.~S. 2012, in Astrophysics and Space Science Library, Vol. 384, Eta
  Carinae and the Supernova Impostors, ed. K.~{Davidson} \& R.~M. {Humphreys},
  221

\end{thebibliography}

\end{document}